\documentclass[floatfix,twocolumn,showpacs,nofootinbib,prd,aps,tightenlines,superscriptaddress]{revtex4-1}
\usepackage{graphicx}
\usepackage{psfrag}
\usepackage{dcolumn}
\usepackage{bm}
\usepackage{amssymb}
\usepackage{amsmath}
\usepackage{amsfonts}
\usepackage{mathrsfs}
\usepackage{bm}
\usepackage{appendix}
\usepackage[caption=false]{subfig}

\newcommand{\bea}{\begin{eqnarray}}
\newcommand{\eea}{\end{eqnarray}}
\newcommand{\beq}{\begin{equation}}
\newcommand{\eeq}{\end{equation}}

\usepackage{color}

\begin{document}

\title{Eccentric Binary Black Hole Simulations with Numerical Relativity}

\author{Giuseppe Ficarra}
\affiliation{Center for Computational Relativity and Gravitation,
School of Mathematical Sciences,
Rochester Institute of Technology, 85 Lomb Memorial Drive, Rochester,
New York 14623, USA}
\affiliation{Dipartimento di Fisica, Universit\`a della Calabria, 
Arcavacata di Rende (CS), 87036, Italy}
\author{Carlos O. Lousto}
\affiliation{Center for Computational Relativity and Gravitation,
School of Mathematical Sciences,
Rochester Institute of Technology, 85 Lomb Memorial Drive, Rochester,
New York 14623, USA}

\date{\today}

\begin{abstract}
We perform a systematic study of eccentric orbiting nonspinning
black hole binaries. We first make a technical study of the optimal
full numerical techniques to apply to these studies.
We find the grid structure and global resolution
that optimize accuracy and speed of current computational resources,
while choosing different gauge parameters and Courant factors,
$c=dx/dt$, find its optimal value of 0.45. 
With these choices we perform a study of the merger times $t_m(e)$
as a function of eccentricity, $e$,
for configurations with comparable orbital energy content and find that
its dependence is well represented by the post-Newtonian factor
$F(e)=(1+73e^2/24+37e^4/96)/(1-e^2)^{7/2}$, when merger times are
normalized to their quasicircular values, i.e. $t_m(e)\approx t_m(0)/F(e)$.
We then perform a systematic coverage of five small-medium eccentricities
up to $e\sim0.45$ and six mass ratios up to 8.5:1 producing a total of 30
simulations covering up to 25 orbits to merger to further model the
unequal mass ratio dependence of merger times and as a seed study to
a forthcoming new systematic catalog of gravitational
waveforms from eccentric binary black holes to allow directly perform parameter
estimations of gravitational waves events.
\end{abstract}

\maketitle

\section{Introduction}\label{sec:Intro}
There is a growing interest in studying highly eccentric binaries, 
where residual eccentricity persists until close to merger. 
These eccentric binaries may produce gravitational wave signals that are particularly intriguing and cannot be accurately modeled using quasicircular approximations 
\cite{Gayathri:2020coq,Romero-Shaw:2022xko,OShea:2021faf,Gupte:2024jfe,Zeeshan:2024ovp}.
This subject has attracted considerable attention 
\cite{ShapiroKey:2010cnz,DOrazio:2018jnv,Hoang:2019kye}, 
but its detailed modeling remains largely incomplete.
Accurate simulations that account for these effects within the 
sensitivity bands of LISA and third generation detectors
can also help to exploit multiband observational opportunities
\cite{Sesana:2016ljz,Vitale:2016rfr,Barausse:2016eii,Bonetti:2017lnj,Bonetti:2018tpf}.

Systematic full numerical simulations of eccentric binaries
has been pioneered in \cite{Gold:2012tk}, and more recently in
\cite{Ramos-Buades:2019uvh,Islam:2021mha,Healy:2022wdn,Ferguson:2023vta}.
A recent paper \cite{Boschini:2024scu} compares definitions of
eccentricity \cite{Campanelli:2008nk,Shaikh:2023ypz} 
during the evolution of full numerical simulations.
Long term simulations have also been used
to model gravitational waveforms for the inspiral, merger and ringdown of nonspinning moderately eccentric binary black hole systems
in \cite{Hinder:2017sxy}.
To improve the new modeling~\cite{Albanesi:2022xge,Albertini:2023aol}
of gravitational waves from eccentric long term merging
binary black holes, extending thus the intrinsic parameter space
to an 8-dimensional one, here we will focus on optimizing the design
of full numerical simulations in order to efficiently systematically
explore this eccentricity dependence.

The paper is organized as follows, in section~\ref{sec:FN} we review the full
numerical techniques used throughout this study. We will analyze gains due
to increasing the Courant factor and choices of the shift damping parameter.
We will also investigate the grid structure and its global resolutions to optimize the
computational cost and accuracy of the simulations.
In Sec.~\ref{sec:eBBH} we will explore the dependence of the merger times
and number of orbits on the initial eccentricity starting from equivalent
total energy configurations of nonspinning equal mass black hole binaries.
In Sec.~\ref{sec:eBBHq} we will apply this optimized set ups to produce a first
30 runs catalog of long term simulations and waveforms from nonspinning merging
binary black holes covering sparsely unequal comparable
mass ratios and up to medium eccentricities.
Finally, we will conclude the paper with a discussion section~\ref{sec:Discussion} 
on how to extend this catalog to additional regions of the parameter space
as well as a tighter coverage with yet improved numerical techniques.

\section{Full Numerical Techniques}\label{sec:FN}

In order to perform the full numerical 
simulations of binary black holes we use the LazEv code\cite{Zlochower:2005bj}
which employs 8th order spatial finite differences \cite{Lousto:2007rj}, 4th order
Runge-Kutta time integration, and a reduced \cite{Zlochower:2012fk} Courant factor $(c=dt/dx=1/4)$.

For setting up numerical initial data for binary black holes,
we regularly adopt the puncture
approach~\cite{Brandt97b} along with the {\sc  TwoPunctures}
~\cite{Ansorg:2004ds} code.
In order to locate apparent horizons during numerical evolutions
we use the {\sc AHFinderDirect}~\cite{Thornburg2003:AH-finding} 
and compute horizon masses from its area $A_H$.
Furthermore, we measure the magnitude of the horizon spins 
$S_H$, using the ``isolated horizon'' algorithm \cite{Ashtekar:2004cn}
as implemented in Ref.~\cite{Campanelli:2006fy}.

We also use the {\sc Carpet}~\cite{Schnetter-etal-03b} mesh refinement driver
to pinpoint the evolution of the black holes across the numerical domain.
{\sc Carpet} provides a ``moving boxes'' style of mesh refinement, 
where refined grids of fixed size are arranged about the
coordinate centers of the holes. These grids are then moved 
following the trajectories of the holes during the numerical simulation.

The grid structure of our mesh refinements have a size of the largest
box for typical simulations of $\pm400M$.  The number of points between 0
and 400 on the coarsest grid is XXX in nXXX (i.e. n100 has 100
points).  So, the grid spacing on the coarsest level is 400/XXX.  The
resolution in the wavezone is $100M/$XXX (i.e. n100 has $M/1.00$, n120
has $M/1.2$ and n144 has $M/1.44$) and the rest of the levels is
adjusted globally. For comparable masses and non-spinning black holes, the grid around one of the black holes
($m_1$) is fixed at $\pm0.65M$ in size and is the 9th refinement level.
Therefore the grid spacing at this highest refinement level is 400/XXX/$2^8$. When considering small mass ratio binaries, we progressively add internal grid refinement levels \cite{Lousto:2020tnb}. Here we set units such that $M=m_1^H+m_2^H$ is
the addition of the horizon masses.

For the gauge choices, we follow the ``moving punctures'' approach 
to dynamically determine coordinates during evolution of the system.
Specifically, we use a modified 1+log lapse and a
modified Gamma-driver shift condition~\cite{Campanelli:2005dd, vanMeter:2006vi}
\begin{subequations}
  \label{eq:gauge}
  \begin{align}
    (\partial_t - \beta^i \partial_i) \alpha &= - \alpha^2 f(\alpha) K \; , \\
    \partial_t \beta^a &= \frac{3}{4} \tilde{\Gamma}^a - \eta \beta^a \;,
  \end{align}
\end{subequations}
where typical initial values of the lapse are $\alpha(t=0)=1/(2\psi_{\text{BL}}-1)$, where   
$\psi_{\text{BL}}=1+m_{(+)}/(2r_{(+)})+m_{(-)}/(2r_{(-)})$, and for the shift
$\beta^a(t=0)=0$, with $\eta=2/M$.

\begin{figure}
\includegraphics[angle=0,width=\columnwidth]{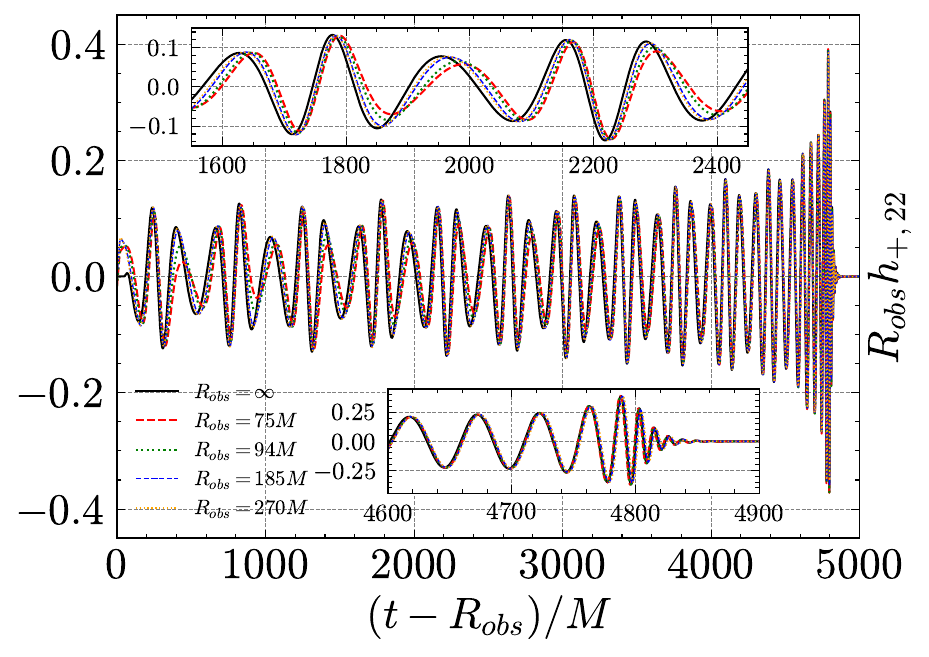}
\caption{$l=m=2$ mode of the strain $h_+$ 
  waveforms as seen at different extraction radii $R_{obs}$,
  and the extrapolated to infinite observer location 
  from the standard choice of $R_{obs}=113M$.
  \label{fig:rex}}
\end{figure}

The extraction of gravitational radiation from the numerical
relativity simulations is performed using the formulas (22) and (23)
from \cite{Campanelli:1998jv} for the energy and linear momentum
radiated, respectively, and the formulas in \cite{Lousto:2007mh}
for angular momentum radiated, all in terms of the extracted Weyl scalar $\Psi_4$
at the observer location $R_{obs}=113M$. In order to extrapolate
the observer location to infinity, we use the perturbative formulas
in Ref.~\cite{Nakano:2015pta}. Fig.~\ref{fig:rex}
provides an illustration of this process of extrapolation in a prototypical
eccentric simulation (EccBBH::03 in Table~\ref{tab:eBBH_initialdata},
with ($e\approx0.24$ eccentricity) for the $l=m=2$ mode of the
strain waveforms $h_+$
at different extraction locations $R_{obs}=75M-270M$ 
compared to the extrapolation to infinity from $R_{obs}=113M$.

To quantify the magnitude of the differences between the waveforms
due to the finite extraction
radii $R_{obs}$ and the extrapolation to infinity,
we use the matching measure,
\begin{eqnarray}
\mathscr{M} \equiv \frac{\left<h_1\left|\right.h_2\right>}{\sqrt{\left<h_1\left|\right.h_1\right>\left<h_2\left|\right.h_2\right>}},
\end{eqnarray}
as implemented via a complex
overlap as described in Eq.~(2) in Ref.~\cite{Cho:2012ed}:
\begin{eqnarray}
\left< h_1 \left|\right. h_2 \right> & = & 2 \int_{-\infty}^{\infty} \frac{d\omega}{S_n(\omega)}\left[\tilde{h}_1(\omega) \tilde{h}_2(\omega)^* \right],\label{eq:match}
\end{eqnarray} where $\tilde{h}(\omega)$ is the Fourier transform of $h(t)$ and $S_n(\omega)$ is the power spectral density of the detector noise (here,
taken to be identically equal
to one since we are interested in the direct waveforms comparisons).
We adopt the leading modes $(\ell,m)=(2,2)$ of $\psi_4$ for the computations
and we do not maximize over an overall constant time shift and
an overall constant phase shift (as is usually done for parameter estimation
of gravitational waves signals~\cite{Lovelace:2016uwp}).

The results are presented in Table~\ref{tab:overlap}. We observe a
monotonic increase in the overlap between the larger extraction radii
waveform and that extrapolated to infinity (from the $R_{obs}=113M$ waveform).
We note here that from the first extraction radius, at  $R_{obs}=75M$
to $R_{obs}=113M$, they lie in
an inner refinement level that is our standard extraction zone,
that from
$R_{obs}=126M$ to $R_{obs}=185M$ extraction radii are on another
external grid refinement level (with half the previous resolution),
and the last extraction radii $R_{obs}=230M$ to $R_{obs}=270M$
at even lower resolutions by a factor 2/3, 
all showing the consistency and accuracy of our extraction method.
In addition, we performed a new simulation with the same binary
parameters and internal refinement levels but extended the two outmost
grid levels from $(\pm200M,\pm400M)$ to $(\pm512M,\pm1024M)$ sizes in order to
extract waveforms at $R_{obs}=320M,460M,496M$ with an spatial
resolution of $2M$ in order to approach the scale of the initial orbital period of the
binary of about $560M$. The results continue to provide an excellent
overlap with the base extrapolated waveform to infinity confirming thus its
accuracy when confronted with extractions directly in the radiation zone.

\begin{table}
\centering
\caption{Matching overlaps at different extraction radii $R_{obs}$ versus the extrapolated to infinite observer location waveform from $R_{obs}/M=113$}
\label{tab:overlap}
\begin{tabular}{ll}
\toprule
$R_{obs}/M$ & Overlap \\
\hline
75.0 &    0.979376059\\
80.0 &    0.983486543\\
87.0 &    0.986744308\\
94.0 &    0.989401237\\
103.0 &   0.991567460\\
{\bf 113.0} &   0.993377831\\
\hline
126.0 &   0.994892071\\
142.0 &   0.996143412\\
145.0 &   0.996332035\\
170.0 &  0.997477586\\
185.0 &  0.997898820\\
\hline
230.0 & 0.998473533 \\
250.0 & 0.998690080 \\
270.0 & 0.998862511 \\
\hline
320.0 & 0.999264914 \\
460.0 & 0.999376578 \\
496.0 & 0.999363088 \\
\hline
\end{tabular}
\end{table}

We have also estimated the overlaps of the extrapolated to infinite
observer locations themselves to show how those final strain waveforms
estimates differ from each other. They are displayed in
Table \ref{tab:matching_extrapolation} where we have also arbitrarily
split the matching of the full waveform into the 'inspiral' part (with
frequencies $M\omega\leq0.1$) and the 'merger' part 
(with frequencies $M\omega\geq0.1$), to verify their accuracy separately.

\begin{table*}
\label{tab:matching_extrapolation}
\caption{Matching overlaps between extrapolated strain waveforms at infinity computed from different extraction radii. 'Inspiral' and 'merger' split is simply given by $M\omega\lessgtr0.1$}
\begin{tabular}{l|ccccccc}
\toprule
($R_{1}/M,R_{2}/M$) & (75,113) & (75,185) & (75,270) & (113,185) & (113,270) & (185,270) & (113,496) \\
\hline
full & 0.9972 & 0.9959 & 0.9956 & 0.9998 & 0.9997 & 0.9999 & 0.9969 \\
inspiral & 0.9491 & 0.9380 & 0.9416 & 0.9982 & 0.9994 & 0.9993 & 0.9884 \\
merger & 0.9999 & 0.9992 & 0.9968 & 0.9993 & 0.9968 & 0.9990 & 0.9994 \\
\hline
\end{tabular}
\end{table*}

The results seem satisfactory for the present parameter
estimation requirements
and precision of the planned full numerical simulations.


\subsection{Designing efficient simulations of eccentric binaries}\label{sec:cfl}

In order to perform our first studies we will consider a
reference simulation starting at $r_+\approx20M$ with ($e\approx0.24$ eccentricity)
initial data labeled as EccBBH::03 in Table~\ref{tab:eBBH_initialdata}
in the Appendix. Then, 
to improve the performance and accuracy of our simulations,
we performed studies involving the time integrations by varying
the Courant factor $c=dt/dx$ relating the spatial and time finite difference
resolutions, as well as choosing optimal global spatial resolutions and
refinement levels to reach acceptable accuracies at the lowest computational
price.

\begin{figure}
\includegraphics[angle=0,width=\columnwidth]{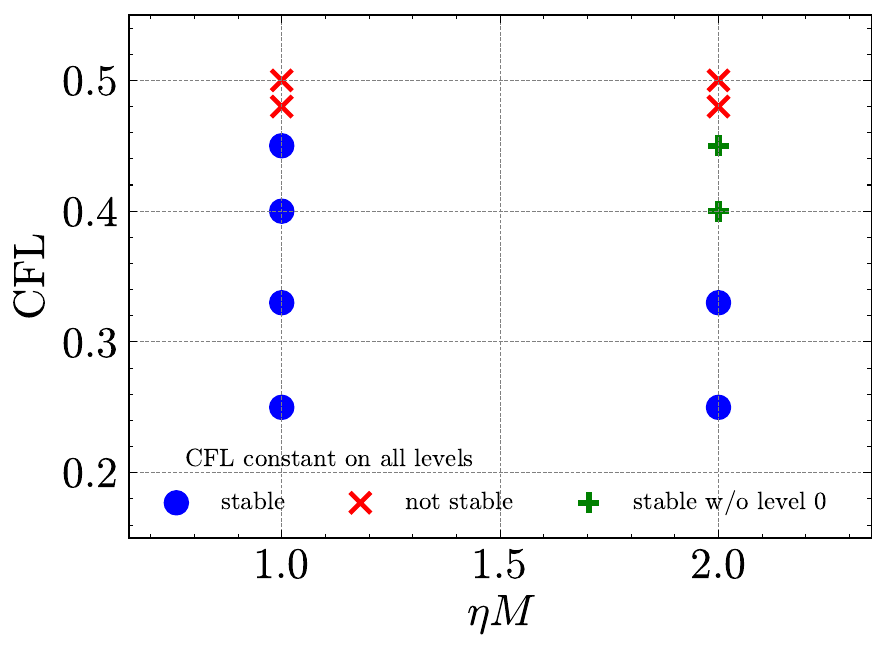}
\caption{Stability map of constant Courant factor $c=dt/dx$ through every grid
  refinement level and for shift evolution parameter $\eta$ in Eq.~(\ref{eq:gauge}).
  \label{fig:cfl}}
\end{figure}

We first design a set of simulations that seek improvements in the speed
of the simulations by increasing the Courant factor from the conservative
$c=1/4$, adopted since Ref.~\cite{Zlochower:2012fk} studies,
to the $c=1/3$ adopted to accelerate the creation of RIT simulations
catalogues \cite{Healy:2017psd,Healy:2019jyf,Healy:2020vre,Healy:2022wdn},
up to the limits of stabilities $c<0.5$. By choosing to keep the
Courant factors constant over all refinements levels to increase the
efficiency of the code, we found the stability limits can be increased
by choosing a lower shift parameter, from $\eta=2/M$ down to $\eta=1/M$ 
as displayed in Figure~\ref{fig:cfl}.
We also note here that we have found benefits in the use of the $\eta=1/M$
gauge when dealing with a quasilocal measure of the black holes recoil
\cite{Healy:2020iuc} and small mass ratio binaries
\cite{Rosato:2021jsq,Lousto:2022hoq}.
Finally, we mention that when we remove the outermost
refinement level for $\eta=2/M$
(bringing effectively the boundary at 200 or alternatively
keeping at 400 with double resolution)
we retrieve the stability limit of $\eta=1/M$, thus showing
that the the instability is
due to underesolution of the outermost simulation grid.
These results are consistent with the studies in
 \cite{Zlochower:2012fk,Rosato:2021jsq}
and the analytic analysis of \cite{Schnetter:2010cz}.

\begin{figure}
  \includegraphics[angle=0,width=\columnwidth]{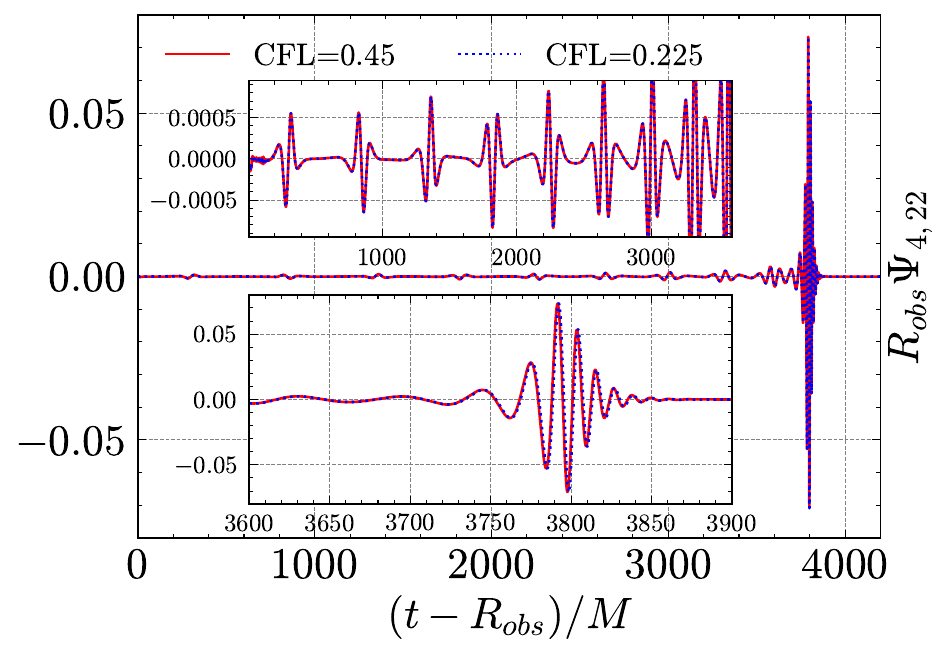}	
  \includegraphics[angle=0,width=\columnwidth]{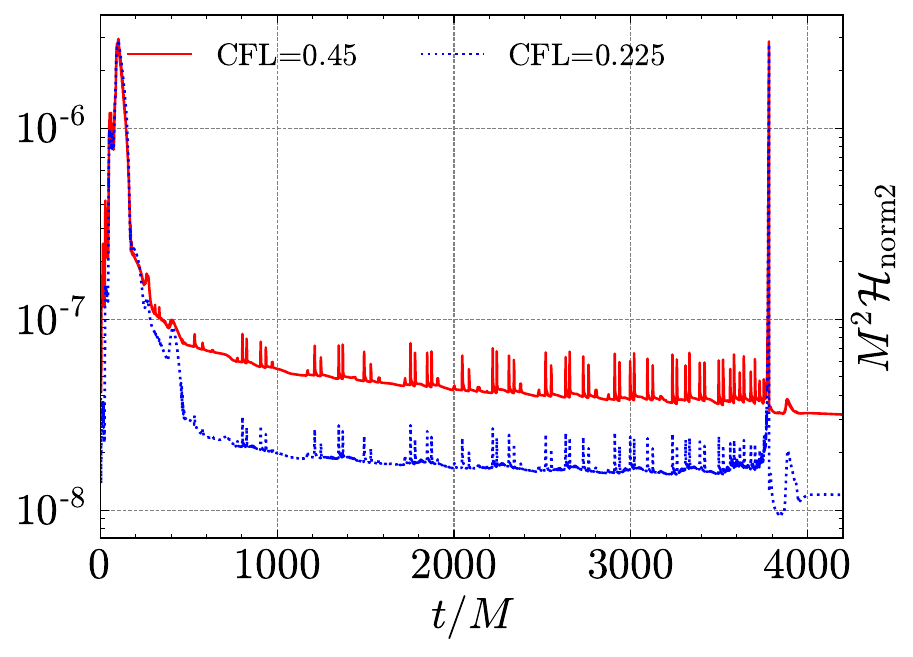}
  \caption{$l=m=2$ mode waveform comparison for a prototypical eccentric binary merger
(EccBBH::03)	   by halving the Courant factor. In the lower panel the effects 
	   on the L2-norm of the Hamiltonian violations.
  \label{fig:cfl2}}
\end{figure}

We also note that for the typical grids and resolutions used in our simulations,
including highly eccentric ones, we observe relatively small differences between
the largest stable Courant factor $c=0.45$ and halving it $c=0.225$ (even lower
than the conservative $c=0.25$). We display the effects of this variation of
the Courant factor in Fig.~\ref{fig:cfl2}.

We will hence assume the Courant factor $c=0.45$ and shift damping parameter
$\eta=1/M$ for the rest of the paper. We next focus on the global resolution
of the grids to attain both, an efficient and accurate set of simulations
with the initial goal to produce a systematic coverage of the parameter space 
of eccentric nonspinning black hole binaries, 
and next to be extended to nonprecessing spinning binaries.

In order to start our eccentric binaries simulations we choose the position
of the apastron $r_+$ and parametrize the tangential initial orbital
momenta by a fraction $1-f$ of the circular linear momentum $P_c$ as
$P_t=P_c(1-f)$\cite{Healy:2022wdn,Ciarfella:2022hfy}.
We thus consider a reference nonspinning equal mass binary with initial
eccentricity defined by $f=0.1$ (corresponding to $e=0.24$)
at an initial separation of $r_+=20M$ and a grid with 8 refinement
levels around each hole.

\begin{figure}
  \includegraphics[angle=0,width=\columnwidth]{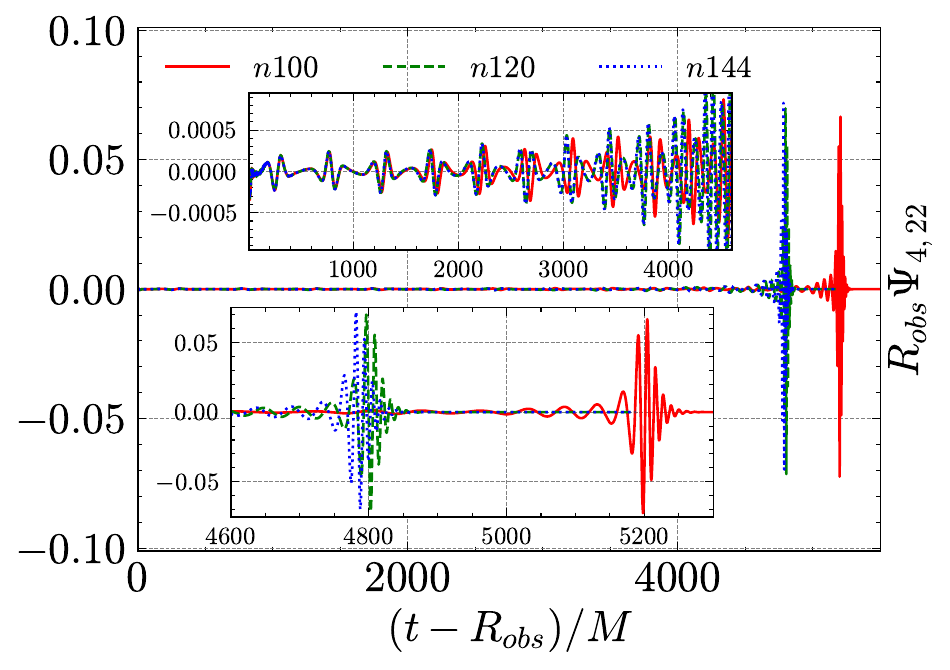}
  \includegraphics[angle=0,width=\columnwidth]{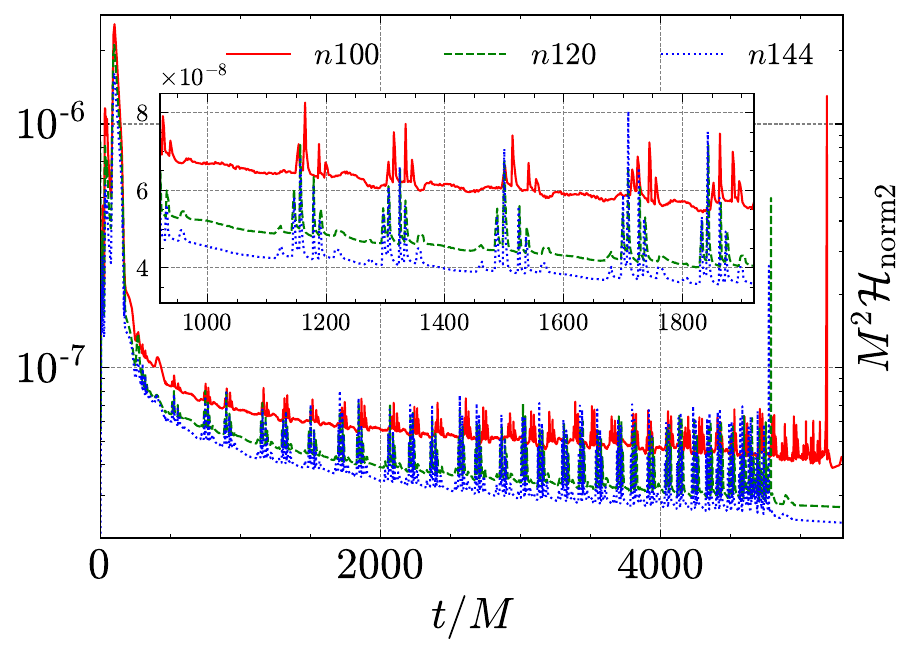}
  \caption{Global convergence study of an eccentric $e=0.24$ equal mass nonspinning binary
    starting from $r_+=20M$.
  \label{fig:resolutions}}
\end{figure}

\begin{figure}
\includegraphics[angle=0,width=\columnwidth]{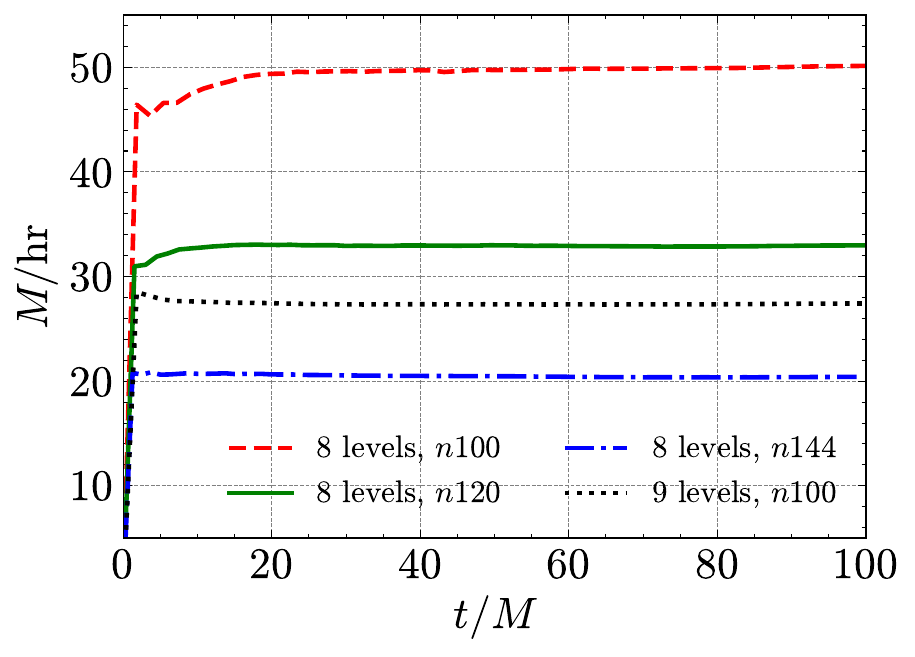}
  \caption{Timing of simulations on 2 Skylake Xeon 6132 compute nodes. 
  \label{fig:speed}}
\end{figure}

The results of the evolutions with increasing global resolutions
by 1.2 factors labeled as n100, n120,
n144, is summarized in Fig.~\ref{fig:resolutions} for waveforms and L2-norms
of the Hamiltonian violations, both displaying strong convergence rates.
We observe the rapid convergence of the merger times towards $4800M$
for n120 and n144, which indicates that n120 provides an accurate enough
resolution without needing to incur into the more (twice) computationally
expensive n144 simulation. 
The alternative resolution n100 with 9 refinement levels
structure used in previous papers as our preferred standard, leads for instance
to a merger time of 4871.9 and a remnant mass of 0.95155345 which improves
over n100 with 8 refinement levels but not over n120 with 8 refinement levels,
according to Table~\ref{tab:eBBH_convergence} results.
This preference is indeed corroborated regarding computational efficiency
directly by observing the timing of the simulations
presented in Fig.~\ref{fig:speed}, performed on 2 compute nodes,
each holding 2 Skylake Xeon 6132 CPUs with 14 cores at 2.6GHz and 192GB of RAM. 
Here we also considered an additional refinement
level around the holes at the lowest resolution n100. This again favors the
choice of the 8 refinement levels and n120 resolution to become
our reference choice for equal mass nonspinning binaries.

  
\begin{table*}
\caption{Convergence of merger time $t_{\rm{m}}/M$, number of orbits $N$, remnant mass $M_f$ and spin $\chi_f$, strain peak amplitude $\left(r/M\right)|h_{22}^{\rm{peak}}|$, peak frequency $M\omega_{22}^{\rm{peak}}$ and peak luminosity $\mathcal{L}_{\rm{peak}}$. Richardson extrapolation is used to determine convergence order and extrapolated values at infinite resolution.\label{tab:eBBH_convergence}
}
\begin{tabular}{lccccccc}
\toprule
resolution & $t_{\rm{m}}/M$ & $N$ & $M_f/M$ & $\chi_f$ & $\left(r/M\right)|h_{22}^{\rm{peak}}|$ & $M\omega_{22}^{\rm{peak}}$ & $\mathcal{L}_{\rm{peak}}$[$10^{-56}$ erg/s] \\
\hline
$n$100 & 5186.70 & 18.93 & 0.951749 & 0.68570 & 0.391546 & 0.357027 & 3.6026 \\
$n$120 & 4787.63 & 17.77 & 0.952027 & 0.68675 & 0.393045 & 0.358207 & 3.6634 \\
$n$144 & 4771.56 & 17.73 & 0.952045 & 0.68640 & 0.392729 & 0.358488 & 3.6783 \\
Inf. Extrap. & 4770.89 & 17.72 & 0.952046 & 0.68649 & 0.392784 & 0.358577 & 3.6830 \\
\hline
Inf.$-$ $n$120 & -16.74 & -0.05 & 0.000019 & -0.00026 & -0.000261 & 0.000370 & 1.9606 \\
$\%$ difference & 0.35 & 0.28 & 0.002 & 0.038 & 0.066 & 0.10 & 0.53 \\
\hline
Conv. order & 17.62 & 17.71 & 15.07 & 5.99 & 8.54 & 7.86 & 7.74 \\
\hline 
\end{tabular}
\end{table*}

\section{eccentric merging binary black holes}\label{sec:eBBH}

With the results of these studies we are ready to select an efficient and
accurate numerical setup to perform a series of simulations of
black hole binaries, to explore its dependence on eccentricity,
as well as on its mass ratios.
In the following, we will choose the base global resolution as n120
and 8 refinement levels for equal mass binaries with corresponding
increase in the refinement levels on the smaller hole
as we consider unequal masses.
In all cases we will choose the maximum Courant factor, 0.45, 
for the gauge with parameter $\eta=1/M$.

\subsection{Equal mass highly eccentric binary black holes evolutions}\label{sec:eBBH1}

When we performed simulations of eccentric binaries from a given initial separation
corresponding to the apastron \cite{Ciarfella:2022hfy},
we noticed a strong drop in the number of orbits and time to merger \cite{Healy:2022wdn}
with increasing eccentricity. However, since we generate eccentric orbits
by dropping the initial tangential linear momentum $P_t$ by a factor $1-f$ from
the quasicircular one $P_c$ at the apastron, i.e. $P_t=P_c(1-f)$, the orbits
are not energetically equivalent.

While strictly, we can only define a unique eccentricity for Keplerian orbits,
at post-Newtonian level it can be generalized by adopting similar looking formulas.
Here we adopt the form $r=a_r (1-e\cos u)$ that has been carried out recently
up to 4PN order studies~\cite{Cho:2021oai}.

Now, in order to compare simulations with different eccentricities on a more
equal footing, we will consider orbits carrying the same ``energy''. To illustrate
this point, let us see how this can be achieved with Newtonian (or Keplerian)
orbits. The apastron can then be defined as $r_+=a(1+e)$ while the ``energy'' of
the orbit goes like $E=-M/2a$. Hence, in order to keep the energy of the system
fixed for different eccentricities, we need to start at the same values of $a$ rather
than $r_+$.
It turns out that this is also true up to first post-Newtonian order, since
$a$ can be expressed as a function of the energy of an eccentric as
\cite{Cho:2021oai}
\bea\label{eq:ar}
a_r=&&\frac{1}{-2E}\left\{1-\frac{E}{2}(\nu-7)+\right.\\
&&\left.\frac{E^2}{4}\left[(1+10\nu+\nu^2
  +(44\nu-68)/(-2Eh^2)\right]+...\right\},\nonumber
\eea
where $\nu=q/(1+q)^2$ and in the last term we can replace $-2Eh^2=(1-e^2)$
at this 2PN order term
(Here $h$ is given by $h = J/(GM)$ with $J$ being the orbital angular
momentum per unit reduced mass).

This allow us to find the initial separation of our simulations as
$r_+=a_r(1+e)$ for families of fixed $E$ values varying the eccentricity $e$.
Note though that, due to the 2PN eccentricity dependence $\sim1/(1-e^2)$,
Eq.~\eqref{eq:ar} breaks down for high values of $e$,
while for small values goes like $\sim e^2$.
We have thus designed a sequence of nonspinning equal mass simulations with
varying eccentricities parameterized by the tangential factor $f(e)=1-P_t/P_c$,
while keeping the reference $a_r$ constant, as displayed in Table~\ref{tab:q1}.
Here we provide a mapping between the more precise eccentricity estimate
of the full numerical simulations $e$ from the 3.5PN approximation as
computed in \cite{Ciarfella:2022hfy} and the simpler Newtonian relation
$e_N = 2f-f^2$, providing the leading dependence, for comparison.

\begin{table}
\centering
\caption{Initial separation of equal mass binaries for different eccentricities}
\label{tab:q1}
\begin{tabular}{lllll}
\toprule
$f$ & $e_N$ & $e$ & $a_r/M$ & $r_+/M$\\
\hline
0.00 & 0.00 & 0.00 & 16.04418 & 16.04418\\
0.05 &0.0975& 0.1242 &  16.04418 & 18.03701\\
0.10 &0.1900& 0.2393 & 16.04418 & 19.88372\\
0.15 &0.2775& 0.3463 & 16.04418 & 21.60086\\
0.20 &0.3600& 0.4452 & 16.04418 & 23.18787\\
0.25 &0.4375& 0.5350 & 16.04418 & 24.81825\\
0.30 &0.5100& 0.6156 & 16.04418 & 25.92176\\
0.325&0.5444& 0.6509 & 16.04418 & 26.48786\\
0.35 &0.5775& 0.6825 & 16.04418 & 26.99510\\
0.40 &0.6400& 0.7342 & 16.04418 & 27.82454\\
0.50 &0.7500& 0.8007 & 16.04418 & 28.89190\\
0.60 &0.8400& 0.8419 & 16.04418 & 29.55227\\
0.70 &0.9100& 0.8739 & 16.04418 & 30.06557\\
0.80 &0.9600& 0.9041 & 16.04418 & 30.55021\\
0.90 &0.9900& 0.9395 & 16.04418 & 31.11799\\
0.95 &0.9975& 0.9662 & 16.04418 & 31.54632\\
1.00 &1.0000& 1.00 & 16.04418 & 32.08837\\
\hline
1.00 &1.0000& \text{headon} & 17.67086 & 35.34172\\
\hline
\end{tabular}
\end{table}

\begin{figure}
\includegraphics[angle=0,width=\columnwidth]{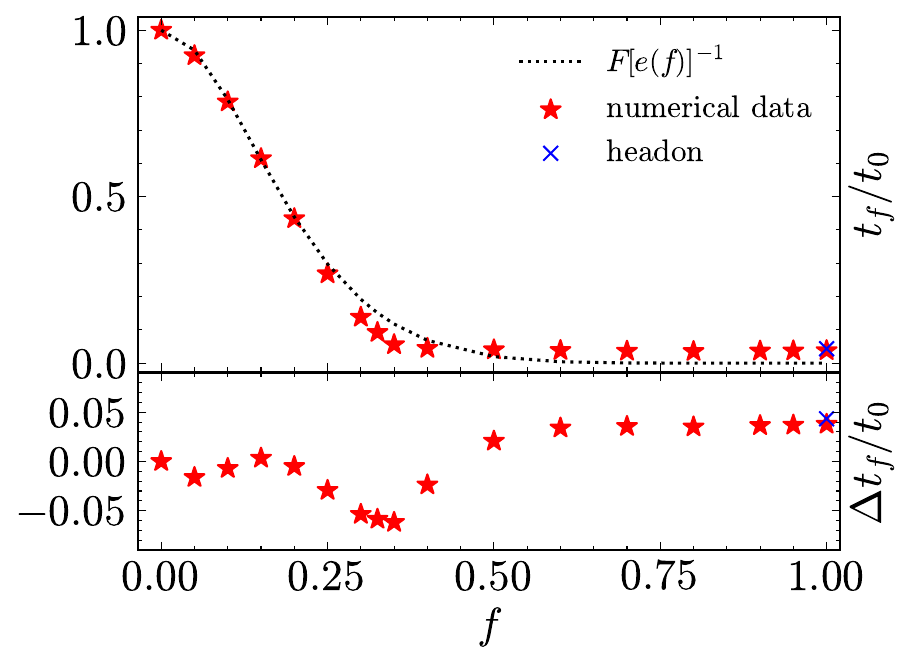}
  \caption{Merger times by the systems with different eccentricity normalized
    to the merger times by the quasicircular reference case as a function of $f$. 
    Bottom panel shows the residuals to the predicted dependence $F(e)$ in Eq.~(\ref{eq:Fe}).
    \label{fig:q1tm}}
\end{figure}

Given that in Eq.~(\ref{eq:ar}) the 2PN eccentricity dependence diverges for $e\to1$, as
a sanity check, we have used explicitly the headon collision case
expression in the Arnowitt-Deser-Misner transverse-traceless
(ADMTT) gauge, from Eq.~(3.18) in Ref.~\cite{Mishra:2010in},
\beq\label{eq:HO}
\epsilon/\mu=E=-\frac{M}{r}\left(1-\frac12\frac{M}{r}+\frac14(1+3\nu)\frac{M^2}{r^2}+...\right),
\eeq
and equated this to the circular 2PN energy in Eq.~(\ref{eq:ar}) with our reference
$a_r=16.04418656$. We find that they match for $r_+=35.34172060$, which gives
an alternative to the one used in our 1PN sequence. This value is displayed at the bottom
of Table~\ref{tab:q1} and in Fig.~\ref{fig:q1tm},
showing the robustness of our approximations and estimates.


The merger times of these simulations $t_f$ as a function of our parametrization
of the eccentricity $f$ are displayed in Fig.~\ref{fig:q1tm}, normalized
to the quasicircular merger time $t_0$. Again, these display the strong decrease of
merger times with increasing eccentricity, this time from nearly equivalent 'energy'
content. Notably, we find that the low-medium eccentricity regime, $f\leq0.25$,
seems to follow the enhancement factor, Eq.~(\ref{eq:Fe}), of \cite{Peters:1963ux},
\beq\label{eq:Fe}
F(e)=\frac{(1+73e^2/24+37e^4/96)}{(1-e^2)^{7/2}}.
\eeq
In drawing these plots, for
the sake of consistency with this low PN order,
we have used the Newtonian relation $e = 2f-f^2$.

The expected leading behavior for the merger time
can be obtained from Eq.~(5.6) of \cite{Peters:1964}
\beq\label{eq:dadt}
<da/dt>=-\frac{64\nu}{5(a/M)^3} F(e),
\eeq
and upon integration over the $a$-dependence and time-dependence.
Since we start all simulations at the same $a=a_r$,
the ratio of the eccentric case to the
circular one ($F(0)=1$) will simply be $t(e)=t(e=0)/F(e)$.
This is closely to what we observe in Fig.~\ref{fig:q1tm},
where we also plot the residuals of the full numerical merger times
to this $F(e)$-dependence normalized to the merger time.
Notably, they lie within a $\sim5\%$ and they can be used 
to display the regime of validity of this approximation,
to small-medium eccentricities reaching up to values of $e\approx0.5$.

\begin{figure}
\includegraphics[angle=0,width=\columnwidth]{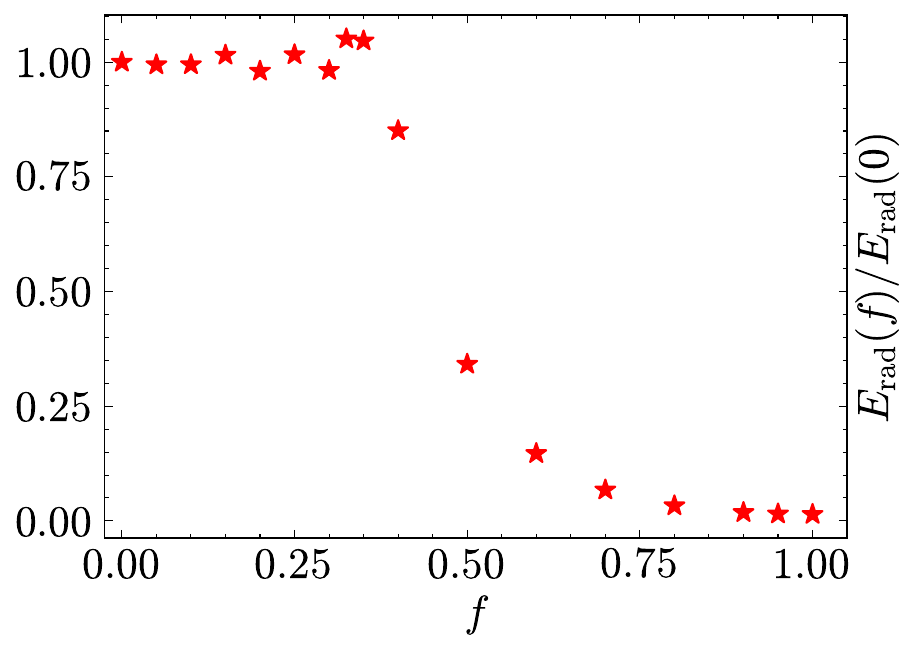}
\caption{Radiated energy by the systems with different eccentricity normalized
  to the energy radiated by the quasicircular reference case as a function of $f$.
  \label{fig:eradf}}
\end{figure}

We can also study the energy radiated versus the eccentricity for our family of
simulations in Table~\ref{tab:q1}. The results are displayed in Fig.~\ref{fig:eradf},
where we observe a certain constancy of the energy radiated versus eccentricity
for, again, low-medium values of $e$, ie. up to values of $f\approx0.35$,
before the values drop due to plunge dominated orbits.

This low-medium $e$ behavior can be found from Eq.~(5.4) of Ref.~\cite{Peters:1964}
\beq\label{eq:dEdt}
<dE/dt>=-\frac{32\nu^2}{5(a/M)^5} F(e),
\eeq
which upon taking the ratio with respect to the circular case and using
that time-dependence of $t(e)=t(e=0)/F(e)$, leads to
$E_{rad}(e)/E_{rad}(0)=F(e)\,t(e)/(F(0)\,t(0))=1$,
which describes roughly the observed constant
behavior for $f\leq0.35$ in Fig.~\ref{fig:eradf}, ie. for small to medium
eccentricities, not directly plunging.

Notably there seems to be a transition from inspiral to direct merger
for intermediate values of $f$ (or $e$). We can explore this further by
taking the ratio of Eqs.~(\ref{eq:dEdt}) and Eq.~(5.7) of Ref.~\cite{Peters:1964}
\beq\label{eq:dedt}
<de/dt>=-\frac{304M^3\nu}{15a^4}e\frac{(1+121e^2/304)}{(1-e^2)^{5/2}},
\eeq
to obtain
\beq\label{eq:dede}
\frac{dE_{rad}}{de}=\frac{6}{19}\frac{M^2\nu}{a}
\frac{(1+73e^2/24+37e^4/96)}{e(1-e^2)(1+121e^2/304)},
\eeq
and to find an inflection point in the behavior $E(e)$ by taking a further
derivative of Eq.~(\ref{fig:eradf}) and equating to zero. We then find 
\beq\label{eq:d2ed2e}
\frac{d^2E_{rad}}{de^2}=0;\quad e_i=0.6054 .
\eeq
Tracing this back to the corresponding Newtonian relation
$f_i=1-\sqrt{1-e_i}$, leads
to $f_i=0.3718$, which notably lies near the transition
observed in Fig.~\ref{fig:eradf}.

It is also interesting to note here that in the radiated
energy displayed in Fig.~\ref{fig:eradf} there seems to be
superposed oscillations of growing amplitude as we approach
the transition values of the eccentricity ($f(e_i)$). These
oscillations are reminiscent of what occurs near critical
phenomena in phase transitions, here separating the orbital
from the direct merger phases. This is beyond the scope of
the current paper and deserves a new exploration in its own
with new targeted fine tuned simulations.

\subsection{Additional studies}\label{sec:q1+}

Here we will explore two additional questions with our eccentric simulations.
One is  regarding the possibility of excitation and amplification of the black hole
quasinormal modes at the closest passage around the periastron. Note that this kind
of phenomena has been observed for the fundamental mode of neutron stars binaries \cite{Arredondo:2021rdt}.
However, black holes are much more stiffer and it was already observed with early simulations
that they, for instance, do not tidal lock at the latest merger stages \cite{Campanelli:2006fg}.

\begin{figure}
  \includegraphics[angle=0,width=\columnwidth]{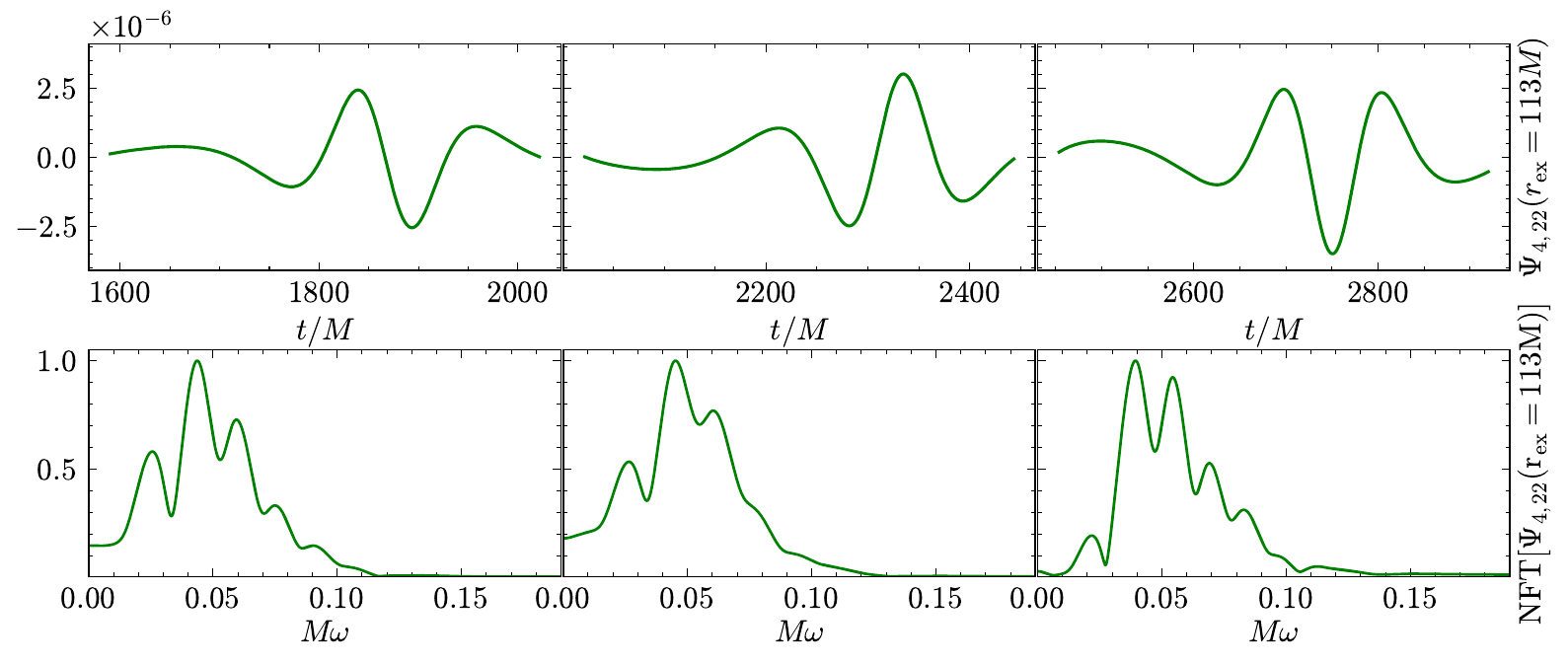}
  \includegraphics[angle=0,width=\columnwidth]{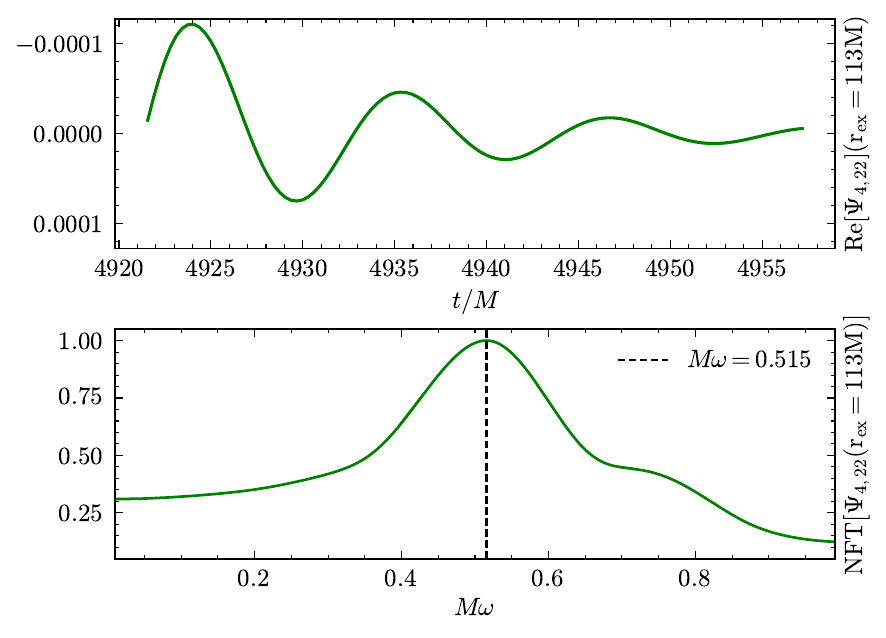}
  \caption{On top the isolated radiation during successive passages through the periastron and its corresponding spectrum. On the bottom the final black hole quasinormal ringing and corresponding spectrum. Note the order of magnitude differences in the characteristic frequencies.
  \label{fig:qnm}}
\end{figure}

In order to visualize this effect in our reference simulations,
we have extracted the
spectrum of radiation of the 4th, 5th, and 6th, closest passages
(out of the total 17 periastron passages down to merger),
from the highest resolution, n144, waveform on top of Fig.~\ref{fig:resolutions}.
In Fig.~\ref{fig:qnm} we show the results of isolating the radiation of those
three successive passages and we see that their spectrum is clearly peaked at
low frequencies, corresponding to a wavelength of the size of the
separation of the closest approach,
thus being dominated by the orbital radiation of gravitational waves, rather than
by the quasinormal modes of the individual holes, that seem to be completely
overshadowed by the orbital radiation. In order to cross check this effect,
we have also computed the spectrum of the final black hole quasinormal ringing,
showing clearly that the characteristic frequency is an order of magnitude
higher than what we observed during the periastron passages.

\begin{figure}
\includegraphics[angle=0,width=0.95\columnwidth]{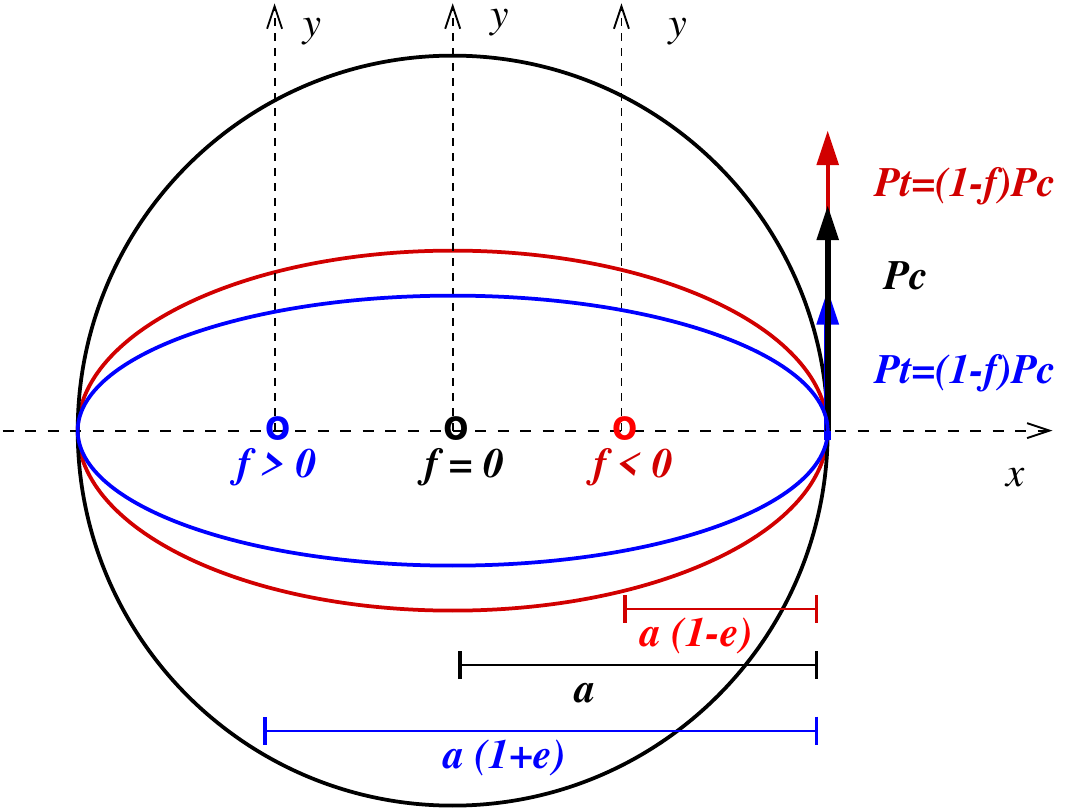}
  \caption{Initial configurations for binaries starting from apastron or periastron.
  \label{fig:forbits}}
\end{figure}

The second observation is that,
while our standard set up to initiate eccentric simulations benefits of the location
of the periastron, it is of interest to generalize it to the case of starting from
the apastron. In that case, our choice of the tangential initial linear momentum
$P_t=(1-f)P_c$, as a fraction of the quasicircular momentum $P_c$ for $0\leq f\leq1$,
can be extended to negative values of $f$ that would lead to eccentric orbits
from the periastron, for sufficiently small negative values, and to hyperbolic
encounters, for larger negative values of $f$. These configurations are depicted
in Fig.~\ref{fig:forbits} where we can observe that starting the simulations with
positive $f\leq1$ as is our standard, one can diminish its values to $f=0$ for the
quasicircular orbit and then move the focus of the ellipse closer to the starting
coordinate of the hole with $f\leq0$, thus transitioning to a periastron initial
configuration.

\begin{figure}
\includegraphics[angle=0,width=\columnwidth]{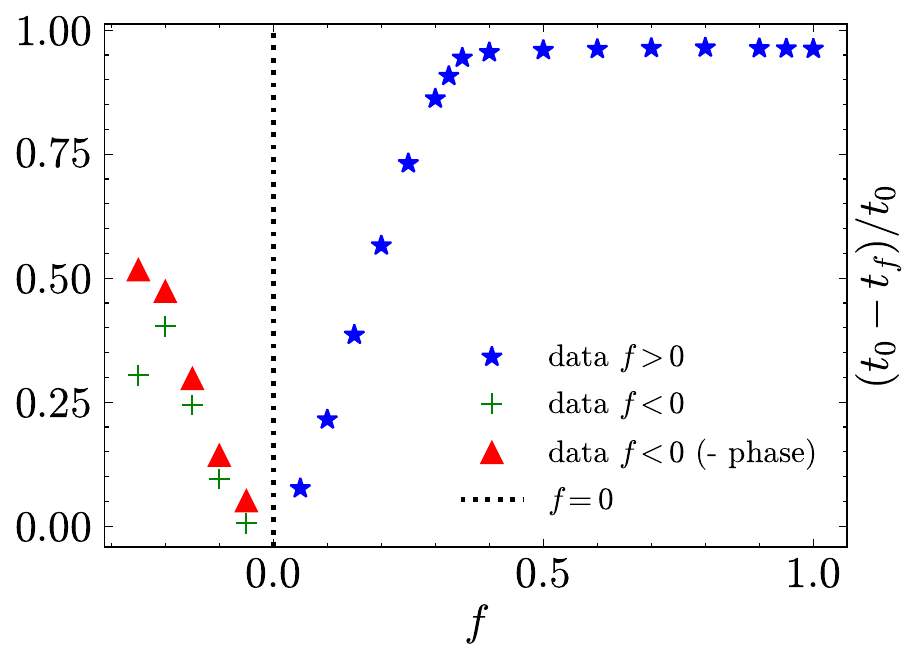}
\caption{Merger times for eccentric orbits with reference to the quasicircular
  case for simulations starting from apastron $f>0$ and periastron $f<0$.
  Correction of the flying time from periastron to apastron (-phase) leads to
  a more symmetric behavior around $f=0$ as expected for small $e$. 
  \label{fig:tmf}}
\end{figure}

We performed five simulations with values of $f=-0.05, -0.2, -0.15, -0.2, -0.25$
and computed the merger times to the quasicircular case as displayed in
Fig.~\ref{fig:tmf}. In order to relate to the corresponding positive values of
$f$ we corrected by the first half orbit flying time from periastron to apastron.
This leads to a more symmetric behavior of the merger time versus $f$ as expected
for small eccentricities,
since once reached the periastron this would put us on an equivalent $-e\to+e$
orbit.
In fact, when we correct for this orbital phase we find a 'V' shape behavior of
the merger times as expected.

We have also verified that the radiated energy for those small $f<0$ cases
remains relatively constant as predicted for $f>0$ and displayed in
Fig.~\ref{fig:eradf}.

As we mentioned, larger negative values of $f$ would lead to hyperbolic orbits.
In order to visualize when roughly this occurs,
one can resort to the Newtonian relationship
$e_N=2f_N-f_N^2$, that leads to the critical
$e=-1$ value for $f_N(-1)=1-\sqrt{2}=-0.4142$. 


\section{unequal mass eccentric binary black holes evolutions}\label{sec:eBBHq}

Here we will perform a series of simulations to systematically cover (sparsely,
for the moment) the parameter space of eccentric, unequal (comparable) nonspinning
black hole binaries. This will serve as a seed for a more thorough coverage of
the (in principle 8-dimensional) parameter space of the binaries.
In doing so we will use the optimized techniques studies presented in Sec.~\ref{sec:cfl}.

To choose the unequal mass configurations we will consider equal spacing,
not in the mass ratio $q=m_1/m_2$ but rather in the symmetric mass ratio
$\nu=q/(1+q)^2$ and restrict to comparable mass ratios $q>1/10$. We will thus
consider the sequence $4\nu=i/8, i=8..3$, thus leading to values of
$q=1, 0.47759225, 1/3, 0.2404082058,$ $0.1715728752, 0.1169631198.$
Taking our base grid of 8 refinement levels for each hole in the equal
mass case, we add one refinement level for the $q=0.47759225$ case around
the smaller hole, and
another one for the $q=1/3, 0.2404082058$ cases, finally reaching 8+11 refinement
levels for the $q=0.1715728752, 0.1169631198$ cases.

The second choice has to do with the fact that starting the simulations
at the same initial separation (or same $a_r$) leads to much larger merger
times simply due to the fact that small mass ratios radiate much less
gravitational energy $\approx\nu^2$. In order to compensate in part for
this effect we consider the leading decay behavior of the orbit in a
post-Newtonian expansion, ie. $\dot r\approx-(64/5)\nu(m/r)^3$ for quasicircular orbits
\cite{Kidder:1995zr}. This leads us to choose the
$a_r(\nu)=a_r(q=1)(i/8)^{1/4}, i=8..3$ for our choice of symmetric mass ratios.
This sequence
leading to the following decreasing values of $a_r$ for decreasing mass ratios,
$a_r(\nu)/M=16.04418656, 15.51742720, 14.93079797,$ $14.26552330, 13.49149896, 12.55525449$.

The combined sequences of 6 mass ratios and 5 eccentricities (parametrized by
the reduction factor $f$) is given in Table~\ref{tab:qeta}, where we have computed
the eccentricity $e$ from $f$ using the 3.5PN formulas in \cite{Ciarfella:2022hfy}
and used the same $a_r$ for a fixed value of each mass ratio $q$.
We did not correct
for the different decay times due to eccentricity by the enhancement factor
\cite{Peters:1963ux,Peters:1964}, since
this will allow us to further verify the $e$-dependence in the unequal mass
cases and extend it to higher order corrections in $\nu$.

\begin{table}
\centering
\caption{Initial separation of unequal mass binaries for different eccentricities}
\label{tab:qeta}
\begin{tabular}{llllll}
\toprule
$4\eta$ & $1/q$ & $a_r/M$ & $f$ & $e$ & $r_+/M$ \\
\hline
8/8 &1.0   & 16.04418 & 0.00  & 0.0 & 16.04418\\
  &    &             & 0.05  & 0.1242    & 18.03701    \\
  &    &             & 0.10  & 0.2393    & 19.88372    \\
  &    &             & 0.15  & 0.3463    &  21.60086   \\
  &    &             & 0.20  & 0.4452   & 23.18787   \\
\hline
7/8 & 2.093836      & 15.51742            & 0.00  & 0.0    &  15.51742  \\
    &  &             & 0.05  & 0.1251    & 17.45910    \\
    &  &             & 0.10  & 0.2410    & 19.25782    \\
    &  &             & 0.15  & 0.3484    & 20.92411    \\
    &  &             & 0.20  & 0.4473    & 22.45870    \\
\hline
6/8 & 3.0   &  14.93079           & 0.00  & 0.0    &   14.93079  \\
    &  &             & 0.05  & 0.1261    & 16.81455    \\
    &  &             & 0.10  & 0.2427    & 18.55584    \\
    &  &             & 0.15  & 0.3506    & 20.16664    \\
    &  &             & 0.20  & 0.4493    & 21.64016    \\
\hline
5/8 & 4.159591      &  14.26552           & 0.00  & 0.0    & 14.26552    \\
    &  &             & 0.05  & 0.1273   & 16.08232    \\
    &  &             & 0.10  & 0.2449    & 17.75931    \\
    &  &             & 0.15  & 0.3530    & 19.30187    \\
    &  &             & 0.20  & 0.4512    & 20.70245    \\
\hline
4/8 & 5.828427      &  13.49149           & 0.00  & 0.0    & 13.49149    \\
    &  &             & 0.05  & 0.1286   & 15.22752    \\
    &  &             & 0.10  & 0.2471    & 16.82567    \\
    &  &             & 0.15  & 0.3554   & 18.28638    \\
    &  &             & 0.20  & 0.4525    & 19.59700    \\
\hline
3/8 & 8.549703      &  12.55525           & 0.00  & 0.0    & 12.55525    \\
    &  &             & 0.05  & 0.1301   & 14.18857    \\
    &  &             & 0.10  & 0.2493    & 15.68615    \\
    &  &             & 0.15  & 0.3572    & 17.04116    \\
    &  &             & 0.20  & 0.4524    & 18.23601    \\
\hline
\end{tabular}
\end{table}

\begin{figure}
\includegraphics[angle=0,width=\columnwidth]{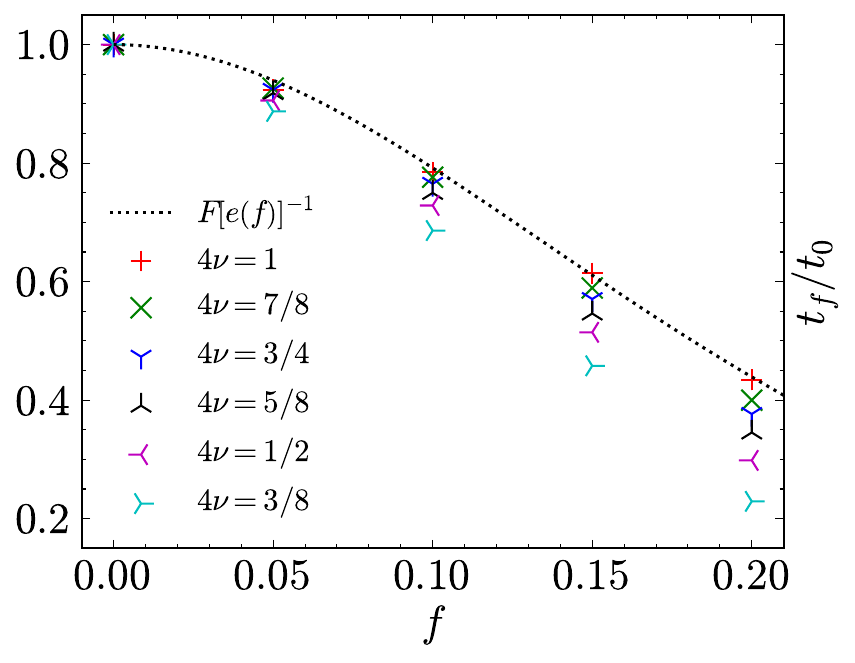}
  \caption{Merger time for unequal mass ratio binaries and residuals to the equal mass cases.
  \label{fig:q_mergertime}}
\end{figure}

Notably, Fig.~\ref{fig:q_mergertime} displays an excellent agreement
between the leading $F(e)$-dependence for equal mass binaries.
We also observe an
increasing residual for decreasing mass ratio binaries and for higher
eccentricities.

In order to model these residuals observed in 
Fig.~\ref{fig:q_mergertime} 
we resource to the studies in Ref.~\cite{Munna:2020som} for the flux
Eq.~(2.1) and in particular the term in Eq.~(2.37),
where a correction to the gravitational radiation flux due to
the mass octupole takes the form
\beq
R_1^{MO}=\frac{(1-4\nu)}{(1-e^2)^{9/2}}\left[
  \frac{1367}{1008}+\frac{18509}{2016}e^2+{\cal O}(e^4)\right],
\eeq
that we will consider as the leading 1PN dependence for our residuals modeling.
We note that the complete 1PN correction, Eq.~(548b)
of Ref.~\cite{Blanchet:2024mnz}
would lead to the same leading dependence for our residuals.

From
\beq
\Delta=\left[dE/dt(e,\nu)-dE/dt(e,4\nu=1)\right]/[dE/dt(e=0,\nu)],
\eeq
that we will use in a low eccentricity expansion to model the unequal mass
correction to the merger time,
\beq
\left[t(e,\nu)-t(e,1/4)\right]/t(0,\nu)\approx\Delta,
\eeq
where
\beq
  \label{eq:fit_modeling}
\Delta=(1-4\nu) (A\,e^2+C) ;\quad e_N^2=(2f_N-f_N^2)^2 .
\eeq
where $A=7703M/504a_r$ and $C=1367M/1008a_r$ for this 1PN octupole correction,
but which we will take them freely as fitting parameters to the full
numerical simulations.

\begin{figure}
\includegraphics[angle=0,width=\columnwidth]{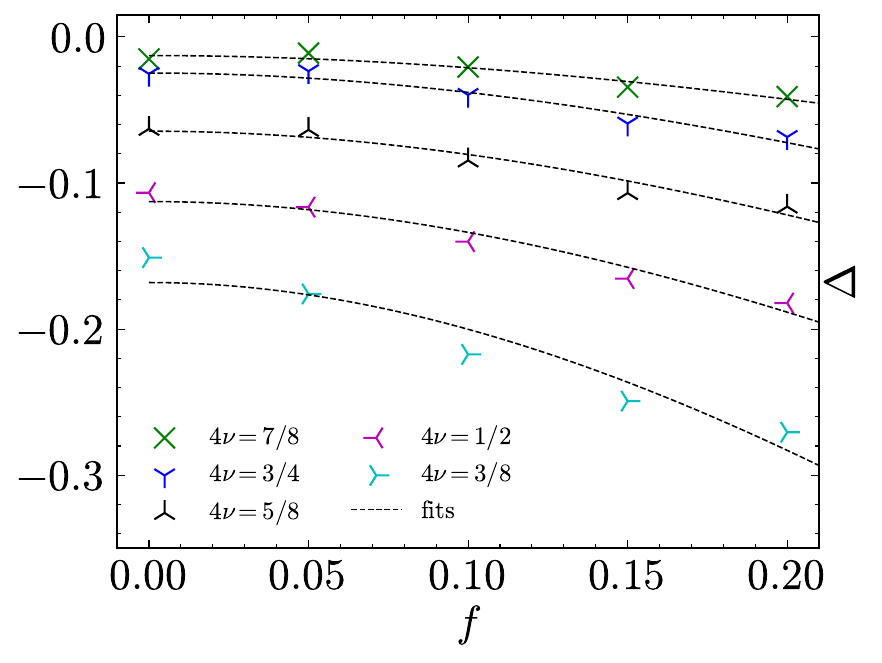}
\caption{Fitting to the merger time correction for unequal mass ratio binaries.
  \label{fig:AC}}
\end{figure}

The results of these fits for each mass ratio sequence are displayed in
Fig.~\ref{fig:AC}.

For the sake of completeness we provide here the fitting values found for these
simulations in Table~\ref{tab:eBBH_qfits} as $A_{N}$ and $C_{NR}$, and where we compare those values with the corresponding 1PN predictions
$A_{1PN}=7703M/504a_r$ and $C_{1PN}=1367M/1008a_r$, finding good agreement. Note that the fitted $C$ also includes residual corrections due to the different $e=0$ starting $a_r$ for different mass ratio simulations.
\begin{table}
\label{tab:eBBH_qfits}
\caption{Fit results for the modeling defined in Eq.~\eqref{eq:fit_modeling}.}
\begin{tabular}{lcccc}
\toprule
$4\nu$ & $A_{NR}$ & $C_{NR}$ & $A_{1PN}$ & $C_{1PN}$\\
\hline
7/8 & $1.849 \pm 0.270$ & $0.013 \pm 0.002$ &0.98494&0.08739\\
6/8 & $1.471 \pm 0.194$ & $0.025 \pm 0.003$ &1.02363&0.09083\\
5/8 & $1.179 \pm 0.175$ & $0.065 \pm 0.005$ &1.07137&0.09506\\
4/8 & $1.168 \pm 0.145$ & $0.113 \pm 0.005$ &1.13284&0.10051\\
3/8 & $1.418 \pm 0.261$ & $0.168 \pm 0.011$ &1.21731&0.10801\\
\hline 
\end{tabular}
\end{table}

Collecting the results of equal and unequal mass binaries, we have an
approximate predictive formula for the merger time $t_m$ of
eccentric to quasicircular orbits of the form
\beq
t_m(e,\nu)/t(e=0,\nu)=1/F(e)+(1-4\nu)(A-\frac{157}{24}C)\,e^2,
\eeq
where
$F(e)=(1+73e^2/24+37e^4/96)/(1-e^2)^{7/2}$, and for the 1PN approximation,
which approximately matches the full numerical values as seen in Table~\ref{tab:eBBH_qfits}, we have $(A-\frac{157}{24}C)=(155125\,M)/(24192\,a_r)$.

\section{Conclusions and Discussion}\label{sec:Discussion}

With the goal of producing a thorough catalog of eccentric binaries,
we have first made a comparative study of full numerical parameters
to optimize the efficiency of such runs in a prototypical case.
We have thus found new options to our standard of simulations to
generate catalogs, including an increase in the Courant factor to
0.45 that can be carried out to all refinement levels provided we
use $M\eta=1$ for the damping parameter of the shift evolution.
We also found that it is more accurate and efficient to trade a
refinement level around black holes for a global 1.2 factor increase
in the resolution of the grids.

We used this optimized full numerical set up to make a first thorough
study of nonspinning equal mass merging binaries versus initial
eccentricity to find that the decrease in the
merger time and number of orbits could
be very well modeled by the leading Post-Newtonian factor 
$F(e)=(1+73e^2/24+37e^4/96)/(1-e^2)^{7/2}$,
when values are normalized by the full numerical
quasicircular, $e=0$, case. Note that the normalization of
$t_m(e)/t_m(e=0)=1/F(e)$ is very important as this is done by
a full numerical simulation to compute $t_m(e=0)$ instead of
using the 1PN estimate, for instance from Eq.~(\ref{eq:dadt}),
thus avoiding the simple prefactor $-(64\nu)/(5(a/M)^3)$
to estimate the merger time of a quasicircular orbit.


The base grid was then scaled to unequal mass ratio binaries by
increasing the number of refinement levels around the smaller hole
\cite{Lousto:2020tnb}, what allowed us to make a first study of the
mass ratio $\nu$-dependence, finding a modeling for the corrections to
the merger times due to unequal masses and small-medium eccentricities.
This in turn can be used to better estimate a priori how to simulate
comparable length waveforms to consistently populate moderately
eccentric binary black hole merger catalogs. For larger eccentricities
it would be interesting to incorporate the modeling of
\cite{Page:2024zpr,Page:2024rsv}.

The success of these simulations and its modeling encourage to produce
further studies of eccentric binaries extending them to spinning
binaries as well as to improve the coverage in length and mass ratios
to be used in parameter estimation studies (as for instance done in
\cite{Healy:2020jjs}) and in model developing.
Those simulations are to be made public in the RIT BBH waveforms
catalog \url{https://ccrgpages.rit.edu/~RITCatalog/}.

\begin{acknowledgments}
The authors thank Manuela Campenelli, Michail Chabanov,
Alessandro Ciarfella, Lorenzo Ennoggi, James Healy,
Hiroyuki Nakano, and Yosef Zlochower for many useful discussions.
The authors also gratefully acknowledge
the National Science Foundation (NSF) for financial support from Grant 
PHY-2207920.  Computational resources were also
provided by the Blue Sky, Green Prairies, and White
Lagoon clusters at the CCRG-Rochester Institute of Technology, which
were supported by NSF grants 
No.\ AST-1028087, No.\ PHY-1229173, No.\ PHY-1726215, and
No.\ PHY-2018420.  This work used the ACCESS allocation TG-PHY060027N,
founded by NSF, and project PHY20007 Frontera, an NSF-funded Petascale
computing system at the Texas Advanced Computing Center (TACC).
\end{acknowledgments}

\appendix
\section{Initial data and results Tables}\label{app:tables}

In Table~\ref{tab:eBBH_initialdata} a small $P_r$ component
has been added to the initial quasicircular parameters from
the instantaneous $dP/dt$ up to 3.5PN radiation terms
\cite{Healy:2017zqj,Ciarfella:2024clj}
in order to ensure a lower eccentricity.
In all other explicit eccentric cases the prescription
consists on setting $P_r=0$ \cite{Ciarfella:2022hfy}.
\begin{table*}
\caption{Initial data parameters for the sequence of simulations described in Sec.\ref{sec:eBBHq} with a larger black hole (labeled 1) and a smaller black hole (labeled 2). Punctures are located at $\Vec{r}_1=(x_1,0,0)$ and $\Vec{r}_2=(x_2,0,0)$ with symmetric mass ratio $\nu$, eccentricity parameter $f$, linear momenta $P=\pm(P_r,P_t,0)$, puncture masses $m^p/M$, horizon (Christodoulou) masses $m^H/M$ and total ADM mass $M_{\rm{ADM}}/M$.\label{tab:eBBH_initialdata}
}
\begin{tabular}{lccccccccccc}
\toprule
Run & $x_1/M$ & $x_2/M$ & $4\nu$ & $f$ & $P_r/M$ & $P_t/M$ & $m^p_1/M$ & $m^p_2/M$ & $m^H_1/M$ & $m^H_2/M$ & $M_{\rm{ADM}}/M$ \\
\hline
EccBBH::01 & 8.02 & -8.02 & 1 & 0.00 & -2.09e-04 & 0.07064 & 0.4912 & 0.4912 & 0.5000 & 0.5000 & 0.9931 \\
EccBBH::02 & 9.00 & -9.00 & 1 & 0.05 & 0.0000 & 0.06253 & 0.4923 & 0.4923 & 0.5000 & 0.5000 & 0.9930 \\
EccBBH::03 & 9.94 & -9.94 & 1 & 0.10 & 0.0000 & 0.05577 & 0.4931 & 0.4931 & 0.5000 & 0.5000 & 0.9930 \\
EccBBH::04 & 10.85 & -10.85 & 1 & 0.15 & 0.0000 & 0.05001 & 0.4938 & 0.4938 & 0.5000 & 0.5000 & 0.9931 \\
EccBBH::05 & 11.68 & -11.68 & 1 & 0.20 & 0.0000 & 0.04506 & 0.4943 & 0.4943 & 0.5000 & 0.5000 & 0.9931 \\
EccBBH::06 & 5.02 & -10.50 & 7/8 & 0.00 & -1.79e-04 & 0.06314 & 0.6691 & 0.3150 & 0.6768 & 0.3232 & 0.9938 \\
EccBBH::07 & 5.64 & -11.82 & 7/8 & 0.05 & 0.0000 & 0.05575 & 0.6701 & 0.3160 & 0.6768 & 0.3232 & 0.9937 \\
EccBBH::08 & 6.22 & -13.04 & 7/8 & 0.10 & 0.0000 & 0.04976 & 0.6707 & 0.3168 & 0.6768 & 0.3232 & 0.9937 \\
EccBBH::09 & 6.76 & -14.17 & 7/8 & 0.15 & 0.0000 & 0.04472 & 0.6713 & 0.3174 & 0.6768 & 0.3232 & 0.9937 \\
EccBBH::10 & 7.25 & -15.21 & 7/8 & 0.20 & 0.0000 & 0.04036 & 0.6717 & 0.3179 & 0.6768 & 0.3232 & 0.9937 \\
EccBBH::11 & 3.74 & -11.19 & 3/4 & 0.00 & -1.50e-04 & 0.05548 & 0.7433 & 0.2426 & 0.7500 & 0.2500 & 0.9945 \\
EccBBH::12 & 4.20 & -12.61 & 3/4 & 0.05 & 0.0000 & 0.04893 & 0.7441 & 0.2435 & 0.7500 & 0.2500 & 0.9945 \\
EccBBH::13 & 4.63 & -13.92 & 3/4 & 0.10 & 0.0000 & 0.04363 & 0.7447 & 0.2442 & 0.7500 & 0.2500 & 0.9944 \\
EccBBH::14 & 5.03 & -15.13 & 3/4 & 0.15 & 0.0000 & 0.03919 & 0.7452 & 0.2448 & 0.7500 & 0.2500 & 0.9944 \\
EccBBH::15 & 5.40 & -16.24 & 3/4 & 0.20 & 0.0000 & 0.03537 & 0.7455 & 0.2452 & 0.7500 & 0.2500 & 0.9944 \\
EccBBH::16 & 2.77 & -11.49 & 5/8 & 0.00 & -1.21e-04 & 0.04762 & 0.8004 & 0.1873 & 0.8062 & 0.1938 & 0.9952 \\
EccBBH::17 & 3.12 & -12.96 & 5/8 & 0.05 & 0.0000 & 0.04193 & 0.8011 & 0.1881 & 0.8062 & 0.1938 & 0.9952 \\
EccBBH::18 & 3.44 & -14.32 & 5/8 & 0.10 & 0.0000 & 0.03736 & 0.8016 & 0.1887 & 0.8062 & 0.1938 & 0.9952 \\
EccBBH::19 & 3.73 & -15.57 & 5/8 & 0.15 & 0.0000 & 0.03354 & 0.8020 & 0.1892 & 0.8062 & 0.1938 & 0.9952 \\
EccBBH::20 & 4.00 & -16.70 & 5/8 & 0.20 & 0.0000 & 0.03027 & 0.8023 & 0.1896 & 0.8062 & 0.1938 & 0.9951 \\
EccBBH::21 & 1.98 & -11.51 & 1/2 & 0.00 & -9.34e-05 & 0.03952 & 0.8487 & 0.1409 & 0.8536 & 0.1464 & 0.9960 \\
EccBBH::22 & 2.23 & -13.00 & 1/2 & 0.05 & 0.0000 & 0.03474 & 0.8493 & 0.1416 & 0.8536 & 0.1464 & 0.9960 \\
EccBBH::23 & 2.46 & -14.37 & 1/2 & 0.10 & 0.0000 & 0.03091 & 0.8497 & 0.1421 & 0.8536 & 0.1464 & 0.9960 \\
EccBBH::24 & 2.67 & -15.62 & 1/2 & 0.15 & 0.0000 & 0.02774 & 0.8501 & 0.1425 & 0.8536 & 0.1464 & 0.9959 \\
EccBBH::25 & 2.86 & -16.74 & 1/2 & 0.20 & 0.0000 & 0.02503 & 0.8503 & 0.1429 & 0.8536 & 0.1464 & 0.9959 \\
EccBBH::26 & 1.32 & -11.23 & 3/8 & 0.00 & -6.71e-05 & 0.03109 & 0.8914 & 0.1002 & 0.8953 & 0.1047 & 0.9968 \\
EccBBH::27 & 1.49 & -12.70 & 3/8 & 0.05 & 0.0000 & 0.02727 & 0.8919 & 0.1008 & 0.8953 & 0.1047 & 0.9968 \\
EccBBH::28 & 1.64 & -14.05 & 3/8 & 0.10 & 0.0000 & 0.02423 & 0.8922 & 0.1012 & 0.8953 & 0.1047 & 0.9968 \\
EccBBH::29 & 1.78 & -15.26 & 3/8 & 0.15 & 0.0000 & 0.02173 & 0.8925 & 0.1016 & 0.8953 & 0.1047 & 0.9968 \\
EccBBH::30 & 1.90 & -16.34 & 3/8 & 0.20 & 0.0000 & 0.01962 & 0.8927 & 0.1018 & 0.8953 & 0.1047 & 0.9967 \\
\hline
\end{tabular}
\end{table*}

In Table~\ref{tab:eBBH_finalproperties} we provide the merged black
hole properties, final mass, spin and recoil velocity of the remnant
hole. We also provide a measure of the radiated gravitational energy,
consistent with the final mass of the merged hole and waveform properties
such as the peak luminosity, amplitude and frequency of the leading
(2,2)-mode. The time and number of orbits to merger of our simulations
completed the table. These properties are given in the format of the
metadata in the RIT BBH waveforms catalog
\url{https://ccrgpages.rit.edu/~RITCatalog/}.
\begin{table*}
\caption{Properties of the sequence of simulations described in Sec.\ref{sec:eBBHq}. We report the remnant mass $M_f/M$ and spin $\chi_f$, the radiated energy $\delta\mathcal{M}=M_{\rm{ADM}}-M_f$, merger time $t_{\rm{m}}/M$, number of orbits $N$, strain peak amplitude $\left(r/M\right)|h_{22}^{\rm{peak}}|$, recoil velocity $V_{\rm{kick}}$, peak frequency $M\omega_{22}^{\rm{peak}}$ and peak luminosity $\mathcal{L}_{\rm{peak}}$.\label{tab:eBBH_finalproperties}
}
\begin{tabular}{lccccccccc}
\toprule
Run & $M_f/M$ & $\chi_f$ & $\delta\mathcal{M}/M$ & $t_{\rm{m}}/M$ & $N$ & $V_{\rm{kick}}$[km/s] & $\left(r/M\right)|h_{22}^{\rm{peak}}|$ & $M\omega_{22}^{\rm{peak}}$ & $\mathcal{L}_{\rm{peak}}$[$10^{-56}$ erg/s] \\
\hline
EccBBH::01 & 0.9516 & 0.6865 & 0.0415 & 6081.0 & 21.37 & 0.000 & 0.3940 & 0.3584 & 3.6818 \\
EccBBH::02 & 0.9518 & 0.6862 & 0.0412 & 5617.5 & 20.14 & 0.000 & 0.3929 & 0.3584 & 3.6611 \\
EccBBH::03 & 0.9520 & 0.6868 & 0.0410 & 4787.6 & 17.77 & 0.000 & 0.3930 & 0.3582 & 3.6634 \\
EccBBH::04 & 0.9509 & 0.6862 & 0.0421 & 3736.1 & 14.58 & 0.000 & 0.3954 & 0.3581 & 3.7005 \\
EccBBH::05 & 0.9524 & 0.6843 & 0.0407 & 2640.0 & 11.21 & 0.000 & 0.3880 & 0.3577 & 3.5628 \\
EccBBH::06 & 0.9623 & 0.6154 & 0.0315 & 5990.4 & 22.00 & 156.11 & 0.3387 & 0.3447 & 2.7009 \\
EccBBH::07 & 0.9625 & 0.6156 & 0.0312 & 5550.6 & 20.66 & 156.72 & 0.3390 & 0.3457 & 2.7037 \\
EccBBH::08 & 0.9626 & 0.6148 & 0.0311 & 4649.6 & 18.01 & 156.72 & 0.3371 & 0.3451 & 2.6762 \\
EccBBH::09 & 0.9624 & 0.6180 & 0.0313 & 3530.1 & 14.64 & 155.43 & 0.3431 & 0.3437 & 2.7604 \\
EccBBH::10 & 0.9618 & 0.6187 & 0.0319 & 2395.3 & 10.83 & 161.50 & 0.3454 & 0.3436 & 2.8224 \\
EccBBH::11 & 0.9712 & 0.5404 & 0.0232 & 5930.8 & 22.45 & 172.65 & 0.2863 & 0.3330 & 1.8967 \\
EccBBH::12 & 0.9713 & 0.5412 & 0.0232 & 5477.8 & 21.35 & 172.74 & 0.2875 & 0.3336 & 1.9293 \\
EccBBH::13 & 0.9715 & 0.5399 & 0.0230 & 4537.4 & 18.35 & 173.35 & 0.2847 & 0.3332 & 1.8748 \\
EccBBH::14 & 0.9717 & 0.5412 & 0.0227 & 3384.2 & 14.60 & 173.73 & 0.2848 & 0.3324 & 1.8928 \\
EccBBH::15 & 0.9719 & 0.5418 & 0.0225 & 2233.6 & 10.61 & 173.57 & 0.2849 & 0.3315 & 1.8962 \\
EccBBH::16 & 0.9787 & 0.4619 & 0.0165 & 5720.9 & 22.92 & 153.72 & 0.2341 & 0.3175 & 1.2775 \\
EccBBH::17 & 0.9789 & 0.4622 & 0.0163 & 5253.0 & 21.36 & 152.10 & 0.2355 & 0.3193 & 1.2672 \\
EccBBH::18 & 0.9786 & 0.4613 & 0.0166 & 4289.2 & 18.14 & 149.66 & 0.2348 & 0.3190 & 1.2563 \\
EccBBH::19 & 0.9784 & 0.4617 & 0.0167 & 3124.5 & 14.30 & 147.80 & 0.2359 & 0.3217 & 1.2828 \\
EccBBH::20 & 0.9784 & 0.4649 & 0.0168 & 1975.5 & 9.81 & 163.17 & 0.2395 & 0.3196 & 1.3149 \\
EccBBH::21 & 0.9850 & 0.3794 & 0.0110 & 5495.0 & 23.57 & 119.31 & 0.1855 & 0.3075 & 0.7984 \\
EccBBH::22 & 0.9850 & 0.3787 & 0.0109 & 4977.0 & 21.67 & 117.31 & 0.1849 & 0.3091 & 0.7816 \\
EccBBH::23 & 0.9851 & 0.3808 & 0.0108 & 4002.8 & 18.17 & 117.38 & 0.1873 & 0.3098 & 0.8046 \\
EccBBH::24 & 0.9852 & 0.3788 & 0.0107 & 2826.8 & 14.07 & 113.45 & 0.1838 & 0.3090 & 0.7794 \\
EccBBH::25 & 0.9846 & 0.3797 & 0.0113 & 1639.1 & 8.91 & 108.73 & 0.1870 & 0.3096 & 0.8027 \\
EccBBH::26 & 0.9902 & 0.2924 & 0.0066 & 5283.0 & 24.40 & 78.04 & 0.1373 & 0.2991 & 0.4392 \\
EccBBH::27 & 0.9901 & 0.2922 & 0.0067 & 4687.5 & 22.15 & 76.26 & 0.1374 & 0.2977 & 0.4342 \\
EccBBH::28 & 0.9901 & 0.2924 & 0.0067 & 3625.1 & 17.95 & 75.24 & 0.1376 & 0.2994 & 0.4340 \\
EccBBH::29 & 0.9903 & 0.2924 & 0.0064 & 2419.1 & 13.10 & 72.68 & 0.1368 & 0.2998 & 0.4322 \\
EccBBH::30 & 0.9791 & 0.2976 & 0.0176 & 1210.5 & 7.21 & 86.67 & 0.1431 & 0.2985 & 0.4761 \\
\hline
\end{tabular}
\end{table*}


\clearpage
\bibliographystyle{apsrev4-1}
\bibliography{../../../Bibtex/references.bib}

\end{document}